\documentstyle[tighten,12pt,psfig,prb,aps]{revtex}

\begin{document}
\title{
Molecular dynamic simulation of a homogeneous {\it bcc}  $\rightarrow${\it hcp}  transition}
\author{J.\ R.\  Morris$^*$ and K.\-M.\ Ho}

\address{
Ames Laboratory--U.S. Department of Energy,\\
Department of Physics and Astronomy,\\
Iowa State University, Ames, IA 50011
}

\maketitle

\begin{abstract}

We have performed molecular dynamic simulations of a Martensitic {\it
bcc}$\rightarrow${\it hcp} transformation in a homogeneous system.
The system evolves into three Martensitic variants, sharing a common
nearest neighbor vector along a {\it bcc} $\langle 111 \rangle$
direction, plus an {\it fcc} region.  Nucleation occurs locally,
followed by subsequent growth.  We monitor the time-dependent
scattering $S({\bf q},t)$ during the transformation, and find
anomalous, Brillouin zone-dependent scattering similar to that
observed experimentally in a number of systems above the
transformation temperature.  This scattering is shown to be related to
the elastic strain associated with the transformation, and is not
directly related to the phonon response.

\draft
\end{abstract}

\section{Introduction.}

 The topic of precursor phenomena in martensitic transformations has
been the subject of numerous studies over the past two decades.
Martensitic systems undergo first-order transitions which are
characterized by a lattice strain (sometimes accompanied by a
``shuffling'' of the atoms associated with a particular phonon
mode).\cite{Krum1,Krum2} Experimentally, it has been observed from
various diffraction measurements that streaking of the Bragg spots
occur in the parent phase close to the transformation, indicating an
incipient instability towards the formation of the martensitic
phase.\cite{Lin,Moss,Axe,Keating,Noda} There is also evidence that
interesting anomalous dynamical behavior occurs concurrently: the
experimentally observed phonon peaks become very broad with enhanced
scattering occurring at low frequencies.\cite{Moss,Axe,Keating,Noda}
More puzzling, the intensities and shapes of these peaks exhibit
unexpected variations between different Brillouin zones. While most of
the experimental data were obtained from alloy systems where the
effects of stoichiometry variations may be of importance, 
similar behavior have also been observed in pure
metals.\cite{PetryLa,Heiming,PetryTi,PetryZr,PetryHf}

In this paper, we explore the relationship between anomalous phonon
behavior and the phase transformation process through molecular
dynamics simulations of  Zr as a prototypical system, exhibiting a {\it bcc} 
to {\it hcp}  transformation. By monitoring
the behavior of the dynamic structure factor of the system,
we examine the microscopic origin of the anomalous phonon behavior in
these systems.  We also study the dynamics of microstructure formation
during the transformation.  To clarify the problem, we will
concentrate on homogeneous systems without defects. The effects of
inhomogeneity on the dynamical behavior of the transition will be left
to a later study.

Scattering lineshapes in first-order structural phase transformations
have been previously studied using various analytic
treatments\cite{Cook,Lindgard} and
simulations\cite{Kerr,Gornostyrev96,Morris92}.  Such
studies, for both first-order and second-order transformations, have
focused on the one-phonon contribution to the scattering, which does
not exhibit a Brillouin zone dependence, even in a full anharmonic
treatment.  Furthermore, most of these
studies\cite{Kerr,Gornostyrev96,Morris92} have focussed on models that
do not exhibit a true elastic transformation: in the high temperature
phase, the linearized phonon frequencies behave as
$\omega^2\rightarrow\omega_0^2+c^2k^2$ as $k\rightarrow 0$, rather
than $\omega\rightarrow ck$.  Such models are more appropriate for
transformations such as $bcc\rightarrow\omega$, characterized by a
``freezing in'' of a phonon mode with a finite frequency (and
wavevector), rather than the development of a static amplitude of an
elastic degree of freedom that occurs in many Martensitic
transformations (including the $bcc\rightarrow hcp$ transformation).

Another issue with previous work examining the dynamic response of
these systems is that most work examines regions of the product (low
temperature) phase embedded in the high temperature
matrix\cite{Cook,Halperin,Bulenda}.  This product phase is
hypothesized to be due to either defects\cite{Halperin,Bulenda}, or
due to long-lived ``heterophase'' fluctuations from the parent phase.
Such studies are not surprising, as long-lived excitations are
necessary to produce ``central peaks'' in the
scattering\cite{Halperin}.  However, in first-order transformations,
the evidence appears to indicate that there is no clear central peak
separate from the phonon peaks (and no long-lived fluctuations into
the product phase), especially in systems with little
disorder.\cite{PetryLa,Heiming,PetryTi,PetryZr,PetryHf,Morris92}

In a previous paper,\cite{Zhang} which we shall refer to as paper I,
we demonstrated that in a homogeneous system, fluctuations in the
high-temperature {\it bcc} phase can produce ``anomalous'' scattering,
with significantly different scattering in equivalent Brillouin zones.
Neutron scattering reveals different line shapes between different
Brillouin zones.\cite{Axe,Noda,PetryLa} Thermal diffuse scattering
(energy integrated scattering), observed in both X-ray
studies\cite{Lin} and neutron diffraction experiments
\cite{Moss,Axe,Keating}, is not featureless, as might be expected from
usual arguments concerning multiple phonon scattering,\cite{Ashcroft}
but instead shows characteristics that imply a connection with the
incipient Martensitic transition.  That such characteristic scattering
can occur in the absence of static or long-lived fluctuations into the
product phase is somewhat surprising, yet is borne out by experimental
work.\cite{PetryLa,Heiming,PetryTi,PetryZr,PetryHf} In paper I, we
argued that these are due to fluctuations {\it toward}, but not into,
the low temperature phase.

More recent molecular dynamics
simulations\cite{Pinsook98,Pinsook99,Pinsook00} have examined the
formation of microstructure during a homogeneous {\it
bcc}$\rightarrow${\it hcp} transformation, utilizing a simple EAM
model for zirconium, as well as the effect of stress on the
microstructure. These simulations demonstrate a number of key points
concerning the formation of microstructure.  Of particular importance
relevant to the work presented here, they have demonstrated that {\it
fcc} regions form at the junction of three {\it hcp} variants sharing
a common $\langle11\bar20\rangle$ direction, rather than forming a
triple point between twin boundaries, and that the twin regions
contain basal plane stacking faults that maintain the 60$^\circ$
orientation between {\it hcp} variants.  The transformation is
interpreted to occur via the Nishiyama-Wassermann
mechanism\cite{Nishiyama} (described in Section III below), rather
than the Burgers mechanism.\cite{Burgers}

Here, we present similar simulations, with the goal of examining the
dynamics during the {\it bcc}$\rightarrow${\it hcp} transition, with a
focus on the aspects relevant to the neutron and x-ray scattering
experiments.  Again, we use molecular dynamics as a tool, to examine
the microscopic processes associated with the dynamics.  We examine
the dynamics in real space, to examine the growth process and the
final microstructure.  The crystallography of the resulting twinned
structure corresponds to experimental results, and those of
Pinsook\cite{Pinsook98,Pinsook99,Pinsook00}.  With respect to the
microstructural development, the most significant difference between
the work presented here and that of Ref.\ \onlinecite{Pinsook98} is
the choice of simulation cell orientation: unlike their simulations,
we do not select out one of the {\it bcc} directions to be along one
of our simulation axes, as that favors {\it hcp} variants with
$\langle 11\bar20\rangle$ along this direction.

We also examine the
time-dependent structure factor $S({\bf Q},t)$, which is the
equivalent of time-dependent x-ray scattering during the transition.
By doing this, we directly demonstrate that the asymmetry in the
scattering between different Brillouin zones is strongest during the
transition, and is essentially due to the formation and motion of new
Bragg peaks during the transition.

\section{Details of simulations.}

We performed the simulations using the same EAM potential as in the
previous paper,\cite{Zhang} and described in detail in Ref.\
\onlinecite{MorrisPMA}.  This potential has a stable {\it bcc} phase,
with phonons in good agreement with experimental results.  The
difference between the cohesive energies between the {\it fcc} and the
{\it hcp} structures is small, compared with {\it ab initio} results,
as is common for these types of potentials.  The EAM potentials for Ti
and Zr are prone to this problem, due to the lack of directional
bonding in these descriptions of these materials.  Another important
caveat is that this potential does not accurately reproduce the
$\omega$-phase properties of Zr or Ti: within this potential, the
$\omega$-phase structure is mechanically unstable.  As a result,
dynamics associated with large-amplitude fluctuations towards this
structure (corresponding to the {\it bcc} {\bf q}=$\frac23[111]$
longitudinal phonon) are not accurately reproduced.  In Zr and Ti (and
especially in certain alloys, including ZrNb), the large amplitude
fluctuations of this will be anharmonic and possibly strongly
non-linear, associated with the corresponding $bcc\rightarrow\omega$
transformation.  This latter transition, while not studied in this
work, is an important topic.\cite{Lin,Moss,Axe,Keating,Noda}  The work
presented here is essentially a model system, examining the
dynamics of the {\it bcc}$\rightarrow${\it hcp} transformation in the
simple case where no competing transformation occurs.

The simulations were performed using two different cubic system sizes:
$18 \times 18 \times 18$ conventional {\it bcc} unit cells (11664
atoms), and $36 \times 36 \times 36$ unit cells (93312 atoms). Except
as noted, all results presented will be those of the larger system.
No qualitative changes occurred as a result of the change in the
system size. Periodic boundary conditions were used in all cases.  In
order to minimally interfere with the dynamics during the transition,
the simulations were performed under constant energy conditions, at an
energy corresponding to the average energy for the {\it bcc} phase at
1500~K.  During the {\it bcc}$\rightarrow${\it hcp} transition, the
system lowered its potential energy, as can be seen in fig.\
\ref{energy}.  With constant energy simulations, this produces a
slight increase in temperature.  For both the $18\times18\times 18$
system and the $36 \times 36 \times 36$ system, this increase was
about 40~K.  This is a minor change in temperature, and we do not
believe that this affects the dynamics of the transition
significantly.  We have also used constant volume simulations, with a
{\it bcc} lattice constant of 3.68 \AA\ as in paper I.  The
transformed structure does not develop any significant macroscopic
stress.  The simulations were performed for 40,000 time steps,
sufficient to observe the transition and subsequent equilibration.
This will be demonstrated in the following section.

\section{Development of microstructure during the transition.}

We begin with a review of the crystallography of the {\it bcc}$\rightarrow$
{\it hcp}  transition.  This is most commonly stated in terms of the
Burgers relationship,\cite{Burgers}
\begin{equation}
(110)_{\it bcc} \| (0001)_{\it hcp} \; \;
 [\bar111]_{\it bcc}\| [11\bar20]_{\it hcp}. \label{Burgers}
\end{equation}
The transformation occurs by ``shuffling'' of alternate $(110)$ planes
in the {\it bcc} lattice, simultaneous with a shearing of the
$(1\bar12)$ planes along the $[\bar111]$ direction.  (In this paper,
we will commonly use three indices for directions and planes in the
{\it bcc} lattice, and four for directions and planes in the {\it hcp}
lattice. We will indicate the lattice structure explicitly where we
wish to emphasize the lattice correspondence, but otherwise will omit
these labels.)  The $(1\bar12)$
direction then becomes a prism plane in the {\it hcp}  lattice.  With these
dynamics in mind, we may rewrite the Burgers relationship
as\cite{Nishiyama}
\begin{equation}
(1\bar12)_{\it bcc} \| (1\bar100)_{\it hcp} \, ; \; \; \;
 [\bar111]_{\it bcc}\| [11\bar20]_{\it hcp}. \label{Nishiyama}
\end{equation}
The shearing of the lattice along $\{1\bar12\}$ planes plays an
important role in understanding the anomalous scattering, as will be
shown in section 4.  Note that if the $[11\bar20]_{\it hcp}$ direction 
is along the $[111]_{\it bcc}$ direction, then there are three
different $\{110\}_{\it bcc}$ planes that can become {\it hcp} basal
planes, namely the $(10\bar1)$, $(01\bar1)$ and $(1\bar10)$ planes.
With four equivalent $\langle111\rangle_{\it bcc}$ directions in the
{\it bcc} lattice, there are 12 {\it hcp} variants that may occur.  In
our simulated microstructure (described below), the system
spontaneously selects out one of the $\langle111\rangle_{\it bcc}$
directions, and forms three {\it hcp} variants from the associated
$\{110\}_{\it bcc}$ planes.

We now establish that there is a transition during our simulations.
The most direct approach is to examine the potential energy vs. time
during the transition.  This is shown in fig.~\ref{energy}, for both
the $18\times18\times18$ and $36 \times 36 \times 36$ system sizes.
As can be seen in the figure, both systems spend considerable amounts
of time (10,000--20,000 time steps) fluctuating about a value near
-6.032 eV.  This corresponds to the value obtained at 1500~K for the
{\it bcc} phase.  Both systems then have a rapid drop in the potential
energy.  This drop continues for some time, then levels off near the
value appropriate for the {\it hcp} system.  Thus, the period during
which the potential energy changes systematically serves to
approximately identify when the transformational dynamics occur.

Before describing the dynamics of microstructure formation, we present
the {\it final} microstructure in the system.  This is shown for the
$36 \times 36 \times 36$ system in fig.\ \ref{schematic36}.  Again,
results for the smaller system are qualitatively the same.  The figure
shows the view down the $[1\bar1\bar1]$ direction.  There are three
different {\it hcp} variants present; the $[1\bar1\bar1]$ direction
forms a $\langle 11\bar20\rangle$ direction in each of these variants,
in accordance with the Burgers relationship.\cite{Burgers} In the
projection shown in fig.\ \ref{schematic36}, the system appears to be
very well ordered.  Each of the variants forms a ``needle-like''
domain, extended indefinitely (via the periodic boundary conditions)
along the $[1\bar1\bar1]$ direction.  By coloring the atoms according
to their potential energy, we have highlighted the twin boundaries
between different {\it hcp} variants.  These boundaries form along hcp
$\langle10\bar11\rangle$ planes, in agreement with experimental
observations.  As in previous
simulations,\cite{Pinsook98,Pinsook99,Pinsook00}, some of the variants
have numerous basal plane stacking faults, allowing for the average
angle between {\it hcp} domains to be 60$^\circ$.  However, specific
twin boundaries deviate noticeably from this value.

We note that in the figure, it is possible to travel from variant II
to variant III without passing through such a twin boundary.  In fact,
there is an {\it fcc} region formed where these variants intersect, as
indicated in the schematic in the figure.  This region is stable
throughout the simulation, and is observed in both the $18 \times 18
\times 18$ and $36 \times 36 \times 36$ systems.  It is not required
to minimize the potential energy of the final configuration: a lower
energy configuration would arise from transforming the variant III
regions to variant I, and the {\it fcc} regions to variant II.  This
would eliminate the total amount of twin boundary, and would also
eliminate the unfavorable {\it fcc} region.  We note that the {\it
fcc}-{\it hcp} energy difference is small in this potential, relative
to ab-initio calculations of pure Zr,\cite{MorrisPMA} suggesting that
{\it fcc} formation in Zr may be decreased or eliminated in the
experimental system.  However, some observations of {\it fcc} Ti have
been made in association with the {\it bcc}$\rightarrow${\it hcp}
transformation, as discussed in ref.\ \onlinecite{Nishiyama}.

We now examine the time evolution of the system, as it progresses from
the {\it bcc} to {\it hcp} phases. This is shown for the $18 \times 18
\times 18$ system in fig.~\ref{atom1.110}.  We focus on the smaller
sized system for clarity; the results for the $36\times 36 \times 36$
system are quite similar.  In the figure, we show a slice of the
system perpendicular to the [110] direction.  The atomic positions
have been averaged over 2000 time steps, in order to eliminate the
high frequency atomic motions, and reveal the slower dynamics
associated with the transformation.  The atoms have been colored
according to their time-averaged potential energy.  Again, this
highlights the development of the microstructure.

The first frame of fig.~\ref{atom1.110}, shows the results for time
steps 18,000-20,000.  Before this, there is little difference between
the structure and that of a perfect {\it bcc}  lattice.  This is consistent
with fig.~\ref{energy}, which shows the potential energy of the
$18\times 18 \times 18$ system beginning to decrease near step 20,000.  In the
first frame, the start of the transformation can be seen just left of
the center of the system.  The vertical rows of atoms show
considerable bending around this region, indicating the large
accommodation strain required for the nucleus.  By the second frame
(steps 20,000-22,000) the transforming region is quite clear.  In
steps 22,000-24,000, the transformed region is starting to develop
sharp boundaries, as it extends along the $[111]$ direction.  This 
transformation progresses and the boundaries develop through step
26,000; by steps 26,000-28,000, the transformation is nearly complete, 
with only the final stages of boundary development remaining.  After
step 30,000 (not shown), the structure is essentially static.

\section{Simulation results for the  structure factor.}

One of our main objectives is to understand the origin of the
pre-martensitic behavior as seen in neutron and x-ray scattering
experiments.  One of the most notable anomalous behaviors is the
difference in thermal diffuse scattering between the so-called
``$\omega$'' points, with strong scattering located at ${\bf Q} =
\frac 13\langle774\rangle$, and significantly less scattering at ${\bf
Q} = \frac 13\langle558\rangle$.  These correspond to equivalent
points in the Brillouin zone, with equivalent magnitudes.
Furthermore, in a simple one-phonon picture, the dependence of the
scattering on the phonon polarizations are also identical.  Therefore,
from the one-phonon picture, the scattering at these points should be
identical.  These modes correspond to the $\frac 23\langle 111\rangle$
longitudinal phonon, which is the mode that leads to the development
of the $\omega$-phase.  This connection between the anisotropy and
possible precursors is very suggestive. However, as shown in paper I
and elaborated on below, a significant difference between these {\bf
Q} vectors is related to the development of the {\it hcp} phase, and
the dynamics of the transformation to this phase.

We begin by showing the structure factor ${S({\bf Q})}$ calculated for
a smaller system, $12\times 12 \times 12$ unit cells, which does not
transform.  The results for scattering in the $(1\bar10)$ plane are
shown in fig.~\ref{sq}.  This has been calculated in the same manner
as the results in paper I.  This figure shows that within our
simulations, the anomalous behavior is reproduced, with the scattering
near ${\bf Q} = \frac 13\langle774\rangle$ significantly greater than
that near ${\bf Q} = \frac 13\langle558\rangle$.  In our potential,
the $\omega$ phase is higher in potential energy than the {\it bcc}
phase, and is mechanically unstable.  Therefore, the results we show
here can not be attributed to $\omega$ phase fluctuations.  The
anomalous scattering can be seen as a ridge extending from the [222]
Bragg peak down to the $\frac 12[552]$ N-point, where a small peak in
the scattering occurs.

We note that in these simulations of the {\it bcc} phase, there is no
evidence of long-lived fluctuations.  The pair correlation function
$g(r)$ exhibits clear peaks associated with the {\it bcc} phase, but
none associated with the {\it hcp} phase.  However, some short-lived
fluctuations {\it toward} the {\it hcp} phase are observed, and are
correlated with anomalous scattering (as described in paper I).  This
is in accordance with both experiments on the high temperature
phase\cite{PetryLa,Heiming,PetryTi,PetryZr,PetryHf} and
theory.\cite{Morris92} Thus, the anomalous scattering can not be
ascribed to previously postulated\cite{Cook} long-lived fluctuations
into the {\it hcp} phase.

In order to explore the scattering behavior, we examine here the
time-dependent scattering at these ``strong'' and ``weak'' $\omega$
points in the Brillouin zone.  As paper I demonstrated, the difference
in scattering between these two points can not be found within the
one-phonon contribution, and the full scattering function must be
calculated.  Moreover, as there are 12 equivalent wave vectors
corresponding to each of these points, we must follow the dynamics of
each point.

Figure \ref{strong} shows $S({\bf Q},t)$ for ${\bf Q} = \frac
13\langle774\rangle$, for the $18 \times 18 \times 18$ system.
As indicated previously, the transformation occurs essentially between
steps 20,000 and 30,000.  In the figure, we see that for three of the
wave vectors, intense scattering occurs during the transformation.
The scattering is as large as 20 times that of the other wave vectors,
or of the same wave vectors before and after the transformation.  For
the other nine wave vectors, the scattering shows no distinctive
behavior before, during, or after the transformation.  Figure
\ref{weak} shows $S({\bf Q},t)$ for ${\bf Q} = \frac 13\langle558\rangle$.  In this case, there is no
distinctive behavior during the transformation.  For three of the wave 
vectors, the scattering increases by roughly a factor of five once the 
transformation is complete.

We can understand these results by examining the full scattering ${S({\bf Q})}$
in the $(1\bar10)$ scattering plane as a function of time.  Rather
than perform the full calculation, we show the scattering calculated
from the {\em time-averaged} positions, such as those shown in
fig.~\ref{atom1.110}.  We show these calculations in figs.\ \ref{sq14}
and \ref{sq24}.  Note that these figures contain essentially static
information about the microstructure.  Before the transformation is
initiated (steps 14,000-16,000), the structure factor shows the Bragg
peaks expected for the {\it bcc}  lattice, plus some weak scattering near the 
N-point phonons at ${\bf Q}=[\frac 52 \frac 52 1]$ and at ${\bf
Q}=[\frac 52 \frac 52 3]$.  Some smearing of the Bragg peaks along the 
$[11\bar2]$ direction is evident.  In steps 16,000-18,000, the
scattering at the N-points has intensified.  

The most dramatic change to the scattering (and the most important to
understanding the anomalous scattering) is seen in steps
18,000-20,000.  The N-point phonons are now developing into static
Bragg peaks, and the Bragg peaks are now shifted from their {\it bcc} 
lattice positions.  The shearing along the $[11\bar2]$ direction is
evident.  This shear brings scattering towards the ${\bf Q} = \frac 13\langle774\rangle$ scattering
vector, while leaving little scattering in the ${\bf Q} = \frac 13\langle558\rangle$ direction.  This
is maintained through step 26,000 - essentially, until the
transformation is nearly complete. The Bragg peaks now are clearly
arranged in a hexagonal pattern.  Thus, the anomalous large scattering
near ${\bf Q} = \frac 13\langle774\rangle$ is not from phonon scattering from $\frac23[111]$ phonons,
but from essentially static displacements associated with N-point
phonons combined with shears in the $[11\bar2]$ direction.  These are
exactly the displacements needed for the formation of the {\it hcp}  phase.
From step 26,000 on, the primary development is the narrowing of the
peaks along the $[111]$ direction, corresponding to the formation of
long, needle-like domains in this direction as seen in the real space
pictures in fig.\ \ref{atom1.110}.  The Bragg peaks also settle into
their final positions, with little remnant scattering in the ${\bf Q} = \frac 13\langle774\rangle$ or
the ${\bf Q} = \frac 13\langle558\rangle$ directions.

We also note that during the course of the transformation, a Bragg
peak develops near ${\bf Q}=\frac 43[111]$.  This is a result of the
differing {\it hcp} variants (and the {\it fcc} region, as well)
developing nearest neighbor lattice vectors along the [111] direction.
For real systems (i.e. Ti and Zr) the presence of this developing
Bragg peak clouds the issue of interpreting the dynamic behavior of
the high temperature phase at this wave vector, as anomalies can be
associated either with the $bcc\rightarrow hcp$ or the
$bcc\rightarrow\omega$ transformation.

\section{Discussion.}

The main purpose of this paper has been to explore the dynamics of the
{\it bcc}$\rightarrow${\it hcp} transformation, and to relate this to
the anomalous diffuse scattering observed in the {\it bcc} phase. By
examining the time-dependent scattering during the scattering, we have
shown that the Brillouin-zone dependence of the scattering is greatest
during the transformation, and is due to the formation and shearing of
the Bragg peaks as they form the {\it hcp} reciprocal lattice.  The
peaks move through the ${\bf Q} = \frac 13\langle774\rangle$ region of
reciprocal space, creating intense scattering at this wave vector, but
their final position is not at this point.  There is no such movement
through the ${\bf Q} = \frac 13\langle558\rangle$ point; hence the
difference in scattering at these points.  A one-phonon picture, taken
from the {\it bcc} lattice, would predict identical scattering at
these points, even in a fully anharmonic theory.  Previous work has
relied upon this
approximation.\cite{Cook,Lindgard,Kerr,Gornostyrev96,Morris92}  

We may now clearly explain the anomalous diffuse scattering
along the ridge between the ${\bf Q}=[222]$ {\it bcc}  Bragg peak and the
N-point ${\bf Q}=\frac 12 [552]$ position.  This scattering is due to
regions of the {\it bcc}  lattice fluctuating towards the {\it hcp}  lattice,
with the result that intensity develops at the N-point, and shifts
along the $[\bar1\bar12]$ direction.  These are precisely the modes
required to take the {\it bcc}  lattice towards the {\it hcp}  lattice, as in
Eq.\ \ref{Nishiyama}.  This interpretation is consistent with previous
simulations of the {\it bcc}  lattice.\cite{Zhang}

We have also examined the microstructure development during the
transformation.  In the beginning of the simulations, there are four
equivalent $\langle111\rangle$ directions in the {\it bcc} lattice.
(Previous simulations have chosen periodic cells that break the
symmetry between these
directions.\cite{Pinsook98,Pinsook99,Pinsook00})  The final
microstructure corresponds to domains of three different {\it hcp}
variants, sharing a common $\langle 11\bar20\rangle$ direction along
one of the original $\langle111\rangle$ directions.  These domains are
extended along this direction as well (corresponding to
``needle-like'' domains).  In addition, there is a domain in the
metastable {\it fcc} lattice.  The final domain structure does not
minimize the potential energy; a lower energy microstructure with two
{\it hcp} domains would satisfy the constraints imposed by the cubic
simulation cell.  Thus, the domain structure must arise out of the
dynamics of the transformation process.\cite{Bales}

The {\it hcp}  variants are separated by $\{10\bar11\}$ twin boundaries, in
accordance with experimental observations.  These boundaries lie
approximately along the {\it bcc}  $\{112\}$ planes, as shown in fig.\
\ref{schematic36}.  However, in some regions, these boundaries are not
flat, caused by basal plane stacking faults within the {\it hcp}  variants.
Pinsook has discussed the formation of these stacking faults in terms
of ``transformation plasticity'',\cite{Pinsook98}, and argues that these
are necessary for the boundaries to lie parallel to the {\it bcc} 
lattice's $\{112\}$ planes, which intersect at 60$^\circ$ degrees,
instead of the slightly different angles formed by an ideal
$\{10\bar11\}$ {\it hcp}  twin boundary (55.9$^\circ$ for an ideal $c/a$
ration).  However, in our simulations, the boundaries often
significantly deviate from being parallel to the {\it bcc}  $\{112\}$
planes, indicating that more stacking faults occur than would be
indicated by these geometric arguments.  In both our simulations and
those in ref.\ \onlinecite{Pinsook98,Pinsook99,Pinsook00}, the {\it hcp}  basal plane stacking
fault energy is lower than expected for Zr,\cite{Morris2000} which
makes comparison with the experimental system difficult.

The dynamics of the transformation correspond to a localized
nucleation event, followed by growth of the domains.  This is seen
quite clearly in fig.\ \ref{atom1.110}.  Such a picture is
inconsistent with recent simulations of a dislocation-nucleated
{\it bcc}$\rightarrow${\it hcp}  transformation,\cite{Gornostryev} in which a
{\it bcc}  ${\bf b}=\langle 100 \rangle \{011\} $ dislocation nucleates the
transition via a long wavelength distortion that evolves into a set of
twin boundaries.  However, this latter work is dependent on the
particular dislocation; similar work with a ${\bf b}=1/2 \langle 110
\rangle \{1\bar10\}$ dislocation showed different
dynamics.\cite{Kuznetsov} This work is also different from that of
Pinsook and Ackland\cite{Pinsook98}, who observed plate-like geometry
very quickly from a quenched simulation.  In this latter work, the
transition is homogeneous; unlike the work presented here, however,
one of the $\langle 111\rangle_{\it bcc}$ directions is singled out in
the simulation cell.  The effect of this on the dynamics is unclear.
They also observed an {\it fcc} region, but in their simulations, this
was a transient phase that was eliminated during the subsequent
microstructural evolution.  The presence of this region is likely to
be sensitive to the choice of interatomic potential; for the work
presented here (and, to a lesser extent, the potential used in Refs.\
\onlinecite{Pinsook98,Pinsook99,Pinsook00}), the difference between
the cohesive energies of the {\it fcc} and {\it hcp} phases is small
compared to the actual value in Ti and Zr, making the appearance here
more likely.

The present work provides a clear demonstration of the relationship
between the dynamics of the {\it bcc}$\rightarrow${\it hcp}
transformation, and the anomalous scattering observed in the parent
phase.\cite{Lin,Moss,Axe,Keating,Noda} The observed zone-dependence
does not require defects or static transformed (or distorted) regions
in the lattice; this is clear from the simulations presented here and
previously,\cite{Zhang}, and from experimental studies of pure
metals.\cite{PetryLa,Heiming,PetryTi,PetryZr,PetryHf} This is an
important step in understanding the experimental results on the
pre-transition dynamics, as measured through scattering experiments.
Further work is required to understand how these dynamics are affected
by defects, compositional fluctuations in alloy systems, and the
competing ${\it bcc} \rightarrow\omega$ transformation.

\acknowledgments

JRM would like to thank Yuri Gornostyrev and M.\ Katsnelson for useful
discussions.  Part of this work was made possible by the Scalable
Computing Laboratory, which is funded by Iowa State University and
Ames Laboratory.  Ames Laboratory is operated for the U.\ S.\
Department of Energy by Iowa State University under Contract
No. W-7405-Eng-82.  This work was sponsored by the Director for Energy
Research, Office of Basic Energy Sciences, by the High Performance
Computing and Communications Initiative, by the Division of Materials
Sciences, Office of Basic Energy Sciences, U. S. Department of Energy.


\pagebreak

\begin{center}
\begin{figure}
\caption[energy]{ Potential energy versus time step, for the $18\times
18\times 18$ and $36 \times 36 \times 36$ system sizes.  The large
change in the potential energies is due to the {\it
bcc}$\rightarrow${\it hcp} transition.  }
\label{energy}
\end{figure}
\end{center}

\begin{center}
\begin{figure}
\caption[schematic]{ 

Final microstructure of $36\times36\times36$ system, as viewed down
the $[1\bar1\bar1]$ direction, in a completely relaxed (zero force)
configuration. The different atomic shadings indicate the potential
energy of the atoms, with large, dark atoms having the highest energy;
large, gray atoms having the lowest energy; and the smaller, lightest
atoms having intermediate energies.  This highlights the twin
boundaries between {\it hcp} variants.  There are three such variants
in the final microstructure, extending along the $[1\bar1\bar1]$
direction, plus an {\it fcc} region as indicated in the schematic
picture shown in the lower right part of the figure.

}
\label{schematic36}
\end{figure}
\end{center}

\begin{figure}
\caption{
Time-averaged positions of atoms during the steps indicated.  The
colors indicate the time-averaged potential energy, with blue being
the lowest and red the highest.  }
\label{atom1.110}
\end{figure}

\begin{center}
\begin{figure}
\caption[strong]{ The static structure facture ${S({\bf Q})}$ calculated from
simulations of the {\it bcc}  phase, as shown in the $(1\bar10)$ scattering
plane.  The x-axis corresponds to the $[110]$ direction; the y-axis
corresponds to the $[001]$ direction.  }
\label{sq}
\end{figure}
\end{center}

\begin{center}
\begin{figure}
\caption[strong]{
Time-dependent scattering at the 12 different ${\bf Q} = \frac 13\langle774\rangle$ wave vectors.  As
can be seen, those wave vectors whose directions are closest to [111]
exhibit large scattering during the transformation.  Note the
differences in scales between graphs.
}
\label{strong}
\end{figure}
\end{center}

\begin{center}
\begin{figure}
\caption[weak]{ Time-dependent scattering at the 12 different ${\bf Q} = \frac 13\langle558\rangle$
wave vectors.  Unlike the scattering at the ${\bf Q} = \frac
13\langle774\rangle$ wave vectors, there is little change in
scattering during the transformation.  }
\label{weak}
\end{figure}
\end{center}

\begin{figure}
\caption[sq14]{
$S({\bf Q})$ calculated from the time-averaged
structures (as shown in fig.\ \ref{atom1.110}) for different time
steps.  The scattering plane is the same is in fig.\ \ref{sq}.  
}
\label{sq14}
\end{figure}

\begin{center}
\begin{figure}
\caption[weak]{
$S({\bf Q})$ calculated from the time-averaged
structures (as shown in fig.\ \ref{atom1.110}) for different time
steps.  The scattering plane is the same is in fig.\ \ref{sq}.  
}
\label{sq24}
\end{figure}
\end{center}

\end{document}